\documentclass{elsart}

\usepackage{natbib}

\usepackage{amssymb}
\usepackage{graphicx}
\usepackage{lineno}

\linenumbers

\newcommand{\va}{v_{\mathrm{A}}}
\newcommand{\cs}{c_{\mathrm{s}}}

\newcommand{\der}{{\rm d}}

\begin{document}

\begin{frontmatter}



\title{Attenuation of small-amplitude oscillations in a prominence-corona model with a transverse magnetic field}


\author{R. Soler, R. Oliver and J. L. Ballester}
\ead{[roberto.soler,ramon.oliver,joseluis.ballester]@uib.es}

\address{Departament de F\'isica, Universitat de les Illes Balears,
              E-07122, Palma de Mallorca, Spain}

\begin{abstract}
Observations show that small-amplitude prominence oscillations are usually damped after a few periods. This phenomenon has been theoretically investigated in terms of non-ideal magnetoacoustic waves, non-adiabatic effects being the best candidates to explain the damping in the case of slow modes. We study the attenuation of non-adiabatic magnetoacoustic waves in a slab prominence embedded in the coronal medium. We assume an equilibrium configuration with a transverse magnetic field to the slab axis and investigate wave damping by thermal conduction and radiative losses. The magnetohydrodynamic equations are considered in their linearised form and terms representing thermal conduction, radiation and heating are included in the energy equation. The differential equations that govern linear slow and fast modes are numerically solved to obtain the complex oscillatory frequency and the corresponding eigenfunctions. We find that coronal thermal conduction and radiative losses from the prominence plasma reveal as the most relevant damping mechanisms. Both mechanisms govern together the attenuation of hybrid modes, whereas prominence radiation is responsible for the damping of internal modes and coronal conduction essentially dominates the attenuation of external modes. In addition, the energy transfer between the prominence and the corona caused by thermal conduction has a noticeable effect on the wave stability, radiative losses from the prominence plasma being of paramount importance for the thermal stability of fast modes. We conclude that slow modes are efficiently damped, with damping times compatible with observations. On the contrary, fast modes are less attenuated by non-adiabatic effects and their damping times are several orders of magnitude larger than those observed. The presence of the corona causes a decrease of the damping times with respect to those of an isolated prominence slab, but its effect is still insufficient to obtain damping times of the order of the period in the case of fast modes.
\end{abstract}

\begin{keyword}
Sun: oscillations \sep Sun: magnetic fields \sep Sun: corona \sep Sun: prominences
\PACS 52.35.Bj \sep 96.60.P- 
\end{keyword}

\end{frontmatter}

\section{Introduction}
\label{sec:intro}

Solar prominences are large-scale coronal magnetic structures whose material, cooler and denser than the typical coronal medium, is in plasma state. Prominences are supported against gravity by the coronal magnetic field, which also maintains the prominence material thermally isolated from the corona. Small-amplitude oscillations in solar prominences were detected almost 40 years ago \citep{harvey}. These oscillatory motions seem to be of local nature and their velocity amplitude is typically less than 2--3~km~s$^{-1}$. Observations have also allowed to measure a wide range of periods between 30~s \citep{balthasar} and 12~h \citep{foullon}. More recently, some high-resolution observations of prominence oscillations by the Hinode/SOT instrument have been reported \citep{okamoto,berger,ofman}. From the theoretical point of view, the oscillations have been interpreted by means of the magnetoacoustic eigenmodes supported by the prominence body. A recent example is the work by \citet{hinode} in which the observations of \citet{okamoto} are interpreted as fast kink waves.  The reader is referred to \citet{oliverballester02,ballester,banerjee} for extensive reviews of both observational and theoretical studies.

Evidence of the attenuation of small-amplitude prominence oscillations has been reported in some works \citep{molowny99,terradasrad,lin2004}. A typical feature of these observations is that the oscillatory motions disappear after a few periods, hence they are quickly damped by one or several mechanisms. The theoretical investigation of this phenomenon in terms of magnetohydrodynamic (MHD) waves has been broached by some authors by removing the ideal assumption and by including dissipative terms in the basic equations. Non-adiabatic effects appear to be very efficient damping mechanisms and have been investigated with the help of simple prominence models \citep{ballai,carbonell,spatial,terradas}. Nevertheless, other damping mechanisms have been also proposed, like wave leakage \citep{schutgensA,schutgensB,schutgensToth}, dissipation by ion-neutral collisions \citep{forteza} and resonant absorption \citep{arregui}.

In a previous work \citep[][hereafter Paper~I]{soler}, we have studied for the first time the wave attenuation by non-adiabatic effects of a prominence slab embedded in the corona. In that work the magnetic field is parallel to the slab axis and it is found that the corona has no influence on the internal slow modes, but it is of paramount importance to explain the damping of fast modes, which are more attenuated than in simple models that do not consider the coronal medium. Following the path initiated in Paper~I, here we investigate the wave damping due to non-adiabatic mechanisms (radiative losses and thermal conduction) in an equilibrium made of a prominence slab embedded in a coronal medium, but now we consider a magnetic field transverse to the slab axis. This configuration and that studied in Paper~I correspond to limit cases, since measurements with Zeeman and Hanle effects indicate that the magnetic field lines are skewed to the long axis of prominences. On average, the prominence axis and the magnetic field form an angle of about 20~deg. Thus, the skewed case is relegated to a future investigation.

The equilibrium configuration assumed here was analysed in detail by \citet{JR92} and \citet{oliver} in the case of ideal, adiabatic perturbations. The main difference between both works is in the treatment of gravity. \citet{JR92} neglected the effect of gravity and so straight field lines were considered. On the other hand, \citet{oliver} took gravity into account and assumed curved field lines according to the \citet{kippen} model modified to include the surrounding coronal plasma \citep{poland}. Despite this difference, both studies agree in establishing a distinction between different normal modes depending on the dominant medium supporting the oscillation. Hence, internal modes are essentially supported by the prominence slab whereas external modes arise from the presence of the corona. In addition, hybrid (or string) modes appear due to the combined effect of both media.

The investigation of the thermal attenuation of oscillations supported by such equilibrium is unsettled to date and, indeed, this is the main motivation for the present study. However, two works \citep{terradasrad01,terradas} studied the wave damping in an isolated prominence slab. \citet{terradasrad01} considered radiative losses given by the Newtonian law of cooling as damping mechanism and studied the attenuation in the \citet{kippen} and \citet{menzel} prominence models. Subsequently, \citet{terradas} considered a more complete energy equation including optically thin radiation, plasma heating and parallel thermal conduction, and assumed straight field lines since gravity was neglected. The main conclusion of both works is that non-adiabatic mechanisms are only efficient in damping slow modes whereas fast modes remain almost undamped. Nevertheless, in the light of the results of Paper~I, the presence of the coronal medium can have an important repercussion on the wave damping. The investigation of this effect is the main aim of the present work. Therefore, we extend here the work of \citet{terradas} by considering the presence of the corona and neglect the effect of gravity as in \citet{JR92} for simplicity.

This paper is organised as follows. Section~\ref{sec:equilibrium} contains a description of the equilibrium configuration and the basic equations which govern non-adiabatic magnetoacoustic waves. Then, the results of this work are extensively discussed in Sect.~\ref{sec:results}. Finally, our conclusions are given in  Sect.~\ref{sec:conclusions}.


\section{Equilibrium and basic equations}
\label{sec:equilibrium}

\begin{figure}[!htb]
\centering
\includegraphics[width=0.95\columnwidth]{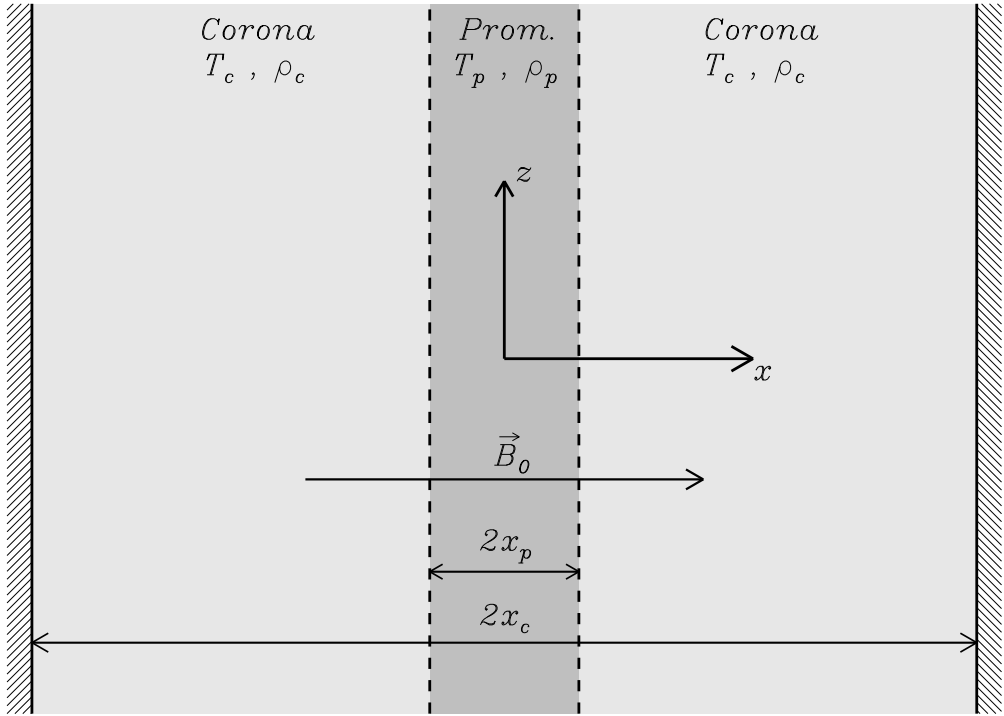}
\caption{Sketch of the equilibrium. The dark region represents the prominence slab while the light region corresponds to the corona. The photospheric walls are the two hatched areas on both sides of the corona.  \label{fig:equilibrium}}
\end{figure}

The equilibrium configuration (see Fig.~\ref{fig:equilibrium}) is made of a homogeneous plasma slab with prominence conditions (density $\rho_{\rm p}$ and temperature $T_{\rm p}$), whose axis is orientated along the $z$-direction, embedded in a coronal environment (density $\rho_{\rm c}$ and temperature $T_{\rm c}$). The system is bounded in the $x$-direction due to the presence of two rigid walls representing the solar photosphere, but it is unlimited in the $y$- and $z$-directions. The width of the prominence slab is $2 x_{\rm p}$ and the total width of the system is $2 x_{\rm c}$. The magnetic field is transverse to the prominence slab, $\vec{B}_0=B_0 \hat{e}_x$, with $B_0$ everywhere constant.

In order to find the basic equations that govern non-adiabatic magnetoacoustic waves we follow the same process as in \citet{terradas}. We consider the usual MHD equations \citep[Eqs.~(1)--(6) of][]{terradas} in which non-adiabatic terms have been included in the energy equation,
\begin{equation}
 \frac{D p}{D t} - \frac{\gamma p}{\rho} \frac{D \rho}{D t} + (\gamma - 1)[\rho L(\rho,T)-\nabla \cdot ({\vec \kappa} \cdot \nabla T)] = 0, \label{eq:eneq}
\end{equation}
where $p$, $\rho$ and $T$ are the gas pressure, density and temperature, respectively. The quantity $\gamma$ is the adiabatic ratio, here taken $\gamma = 5/3$. The non-ideal terms in Eq.~(\ref{eq:eneq}) are explained in detail in \citet{carbonell}. Thermal conduction is represented by $\nabla \cdot ({\vec \kappa} \cdot \nabla T)$, where $\vec \kappa$ is the conductivity tensor which in coronal and prominence applications is usually approximated by its parallel component to the magnetic field,  $\kappa_\parallel = 10^{-11} T^{5/2}\, \mathrm{W}\, \mathrm{m}^{-1}\, \mathrm{K}^{-1}$. Radiative losses and heating are evaluated together through the heat-loss function, $L(\rho,T)= \chi^* \rho T^\alpha - h \rho^a T^b$, where radiation is parametrised with $\chi^*$ and $\alpha$ (see Table~I of Paper~I) and the heating scenario is given by exponents $a$ and $b$ \citep{rosner,dahlburg}.

Regarding our equilibrium configuration, the reader must be aware that, although there have been some attempts  to construct a self-consistent prominence model including both magnetostatics and thermodynamics \citep[e.g.][]{milne,low,anzer}, to date this task remains to be done. Here, we consider a simplified prominence-corona configuration, but it includes the two basic ingredients observed in real prominences. First, the existence of a steep temperature gradient between the prominence and the corona and, second, the apparent thermal isolation of the prominence material from the much hotter corona. The first point is addressed by considering that the temperature profile is a step function, and so the prominence-corona transition region (PCTR) has not been considered. This choice is supported by results of previous works \citep[e.g.][]{oliverballester96} which showed that the PCTR has a minor influence on the prominence oscillatory modes. On the other hand, to represent the thermal isolation we have neglected the heat flux due to thermal conduction at the boundary between the prominence and the corona. Therefore, we impose that both the prominence and the corona are isothermal and thermally isolated, and so radiative losses and heating are locally balanced, i.e. $L(\rho_0,T_0)=0$, where $\rho_0$ and $T_0$ are the local equilibrium density and temperature, respectively.

Assuming that the plasma is at rest in the equilibrium state (i.e. no flux of material) and considering small perturbations, we find the linearised version of the MHD equations \citep[Eqs.~(10)--(15) of][]{terradas}. According to the geometry of our model, we assume perturbations of the form $f_1(x)\exp i (\omega t + k_y y + k_z z)$ and exclude Alfv\'en waves from this analysis by considering only motions and propagation in the $xz$-plane ($v_y=0$, $k_y=0$). Now we combine the resultant expressions and eliminate all the perturbed quantities in favour of the velocity perturbations, $v_x$ and $v_z$, and the temperature perturbation, $T_1$. By this process, we obtain three coupled ordinary differential equations,
\begin{equation}
 \cs^2 \frac{\der^2 v_x}{\der x^2} + \gamma \omega^2 v_x + i k_z \cs^2 \frac{\der v_z}{\der x} - \frac{i \omega \cs^2}{T_0} \frac{\der T_1}{\der x} = 0, \label{eq:basic1}
\end{equation}
\begin{equation}
 \va^2 \frac{\der^2 v_z}{\der x^2} + \left[ \omega^2 - k_z^2 \left( \va^2 + \frac{\cs^2}{\gamma} \right) \right] v_z + i k_z \frac{\cs^2}{\gamma} \frac{\der v_x}{\der x} + \omega k_z \frac{\cs^2}{\gamma} \frac{T_1}{T_0} = 0, \label{eq:basic2}
\end{equation}
\begin{eqnarray}
\kappa_\parallel \frac{1}{p_0} \frac{\der^2 T_1}{\der x^2} &-& \left( \omega_T + \frac{i \omega}{\gamma -1} \right) \frac{T_1}{T_0} \nonumber \\ &-& \left( 1+ \frac{i \omega_\rho}{\omega} \right) \frac{\der v_x}{\der x} - i k_z  \left( 1+ \frac{i \omega_\rho}{\omega} \right) v_z = 0, \label{eq:basic3}
\end{eqnarray}
where $\cs^2 = \frac{\gamma p_0}{\rho_0}$ is the adiabatic sound speed squared whereas $\va^2 = \frac{B_0^2}{\mu \rho_0}$ is the Alfv\'en speed squared. $p_0$ and $B_0$ denote the equilibrium gas pressure and magnetic field strength, respectively, and $\mu$ is the magnetic permittivity ($\mu = 4 \pi 10^{-7}$ in MKS units). Quantities $\omega_T$ and $\omega_\rho$ are defined as follows,
\begin{displaymath}
\omega_\rho \equiv \frac{\rho_0}{p_0}\left( L + \rho_0 L_\rho \right), \qquad \omega_T \equiv \frac{\rho_0}{p_0}T_0 L_T,
\end{displaymath}
$L_\rho$, $L_T$ being the partial derivatives of the heat-loss function with respect to density and temperature, respectively,
\begin{displaymath}
L_\rho \equiv \left( \frac{\partial L}{\partial \rho} \right)_T, \qquad 
L_T \equiv \left( \frac{\partial L}{\partial T} \right)_\rho.
\end{displaymath}

Equations~(\ref{eq:basic1}), (\ref{eq:basic2}) and (\ref{eq:basic3}) govern fast and slow magnetoacoustic waves together with the thermal or condensation mode. In this work we do not study the thermal wave since we pay our attention to the magnetoacoustic modes. \citet{terradas} found an approximate analytical solution of Eqs.~(\ref{eq:basic1})--(\ref{eq:basic3}) by neglecting thermal conduction, a valid assumption in prominence plasmas. However, thermal conduction has an important role in coronal conditions and cannot be neglected in order to perform a realistic description of the oscillatory modes supported by our equilibrium configuration. Hence, we solve the full set of Eqs.~(\ref{eq:basic1})--(\ref{eq:basic3}) using the numerical code PDE2D \citep{sewell} based on finite elements \citep[see][for an explanation of the method]{terradas}. The jump conditions at the interface between the prominence and the corona are automatically well-treated by the code. These jump conditions are \citep{goed}:
\begin{equation}
  \left[ \vec{v} \right] = \vec{0}, \quad \left[ \vec{B} \right] = \vec{0}, \quad \left[ p \right] = 0, 
\end{equation}
where $\vec v$ and $\vec B$ are the perturbed velocity and the magnetic field vectors, respectively. For a complete closure of the system we need to supply a physically consistent set of boundary conditions for the perturbations at the photospheric walls, $x = \pm x_{\rm c}$. In this work, we consider two different sets of boundary conditions,
\begin{equation}
\begin{array}{lll}
 v_x  = v_z  = T_1  = 0, & \mathrm{at} & x = \pm x_{\rm c}, \label{eq:set1}
\end{array}
\end{equation}
and
\begin{equation}
\begin{array}{lll}
  v_x  = v_z = T_1' = 0, & \mathrm{at} & x = \pm x_{\rm c}, \label{eq:set2}
\end{array}
\end{equation}
where $'$ indicates derivative with respect to $x$. Both sets consider line-tied conditions for the velocity perturbations, i.e. the disturbances are unable to perturb the dense photospheric plasma which acts as perfectly rigid wall. On the other hand, sets~(\ref{eq:set1}) and (\ref{eq:set2}) differ by the condition for $T_1$, which has different physical implications. Set~(\ref{eq:set1}) assumes that the perturbation to the temperature vanishes at $x = \pm x_{\rm c}$ and this means that the photospheric walls are taken as isothermal. On the contrary, set~(\ref{eq:set2}) considers a zero-temperature gradient for the perturbation between the corona and the photosphere, so no perturbed heat flux is allowed at the boundaries. From our point of view, set~(\ref{eq:set1}) makes more physical sense than set~(\ref{eq:set2}), since one can expect that the much denser photospheric plasma can instantaneously radiate away any incoming perturbed heat flux from the corona. However, set~(\ref{eq:set2}) imposes that there is no heat exchange between the corona and the photosphere, although the temperature perturbation can have a non-zero value at the walls. Regarding these boundary conditions, \citet{cargill} performed a study of the thermal stability of wave and thermal modes in a Cartesian coronal slab and pointed out that the solutions computed by assuming the boundary conditions given by set~(\ref{eq:set1}) are more thermally stable than those obtained for boundary conditions of set~(\ref{eq:set2}). 

For a fixed real $k_z$, the numerical solution of Eqs.~(\ref{eq:basic1}), (\ref{eq:basic2}) and (\ref{eq:basic3}) provides with a complex frequency, $\omega = \omega_{\rm R} + i \omega_{\rm I}$.  In the ideal, adiabatic case $\omega_{\rm I} = 0$ and therefore the solutions of Eqs.~(\ref{eq:basic1}), (\ref{eq:basic2}) and (\ref{eq:basic3}) are those of \citet{JR92}. Using the real and imaginary parts of the frequency, we can compute the oscillatory period, $P$, the damping time, $\tau_{\rm D}$, and the ratio of both quantities,
\begin{equation}
P = \frac{2 \pi}{\omega_{\rm R}}, \qquad \tau_{\rm D} = \frac{1}{\omega_{\rm I}}, \qquad \frac{\tau_{\rm D}}{P} = \frac{1}{2\pi} \frac{\omega_{\rm R}}{\omega_{\rm I}}.
\end{equation}


\section{Results}
\label{sec:results}

Unless otherwise stated, the following equilibrium parameters are considered in all computations: $T_{\rm p}=8000$~K, \mbox{$\rho_{\rm p} = 5 \times 10^{-11}$~kg~m$^{-3}$}, $T_{\rm c} = 10^6$~K, $\rho_{\rm c} = 2.5 \times 10^{-13}$~kg~m$^{-3}$, $B_0=5$~G, $x_{\rm p} = 3000$~km and $x_{\rm c} = 10 x_{\rm p}$. The coronal density is computed by fixing the coronal temperature and imposing pressure continuity across the interfaces. In addition, we assume an optically thin prominence plasma (regime Prominence~(1) of Paper~I) and a constant heating per unit volume ($a=b=0$). In all the following expressions, subscript  0 indicates local equilibrium values, while subscripts p and c denote quantities explicitly computed with prominence and coronal parameters, respectively. 

\subsection{Dispersion diagram and wave modes}
\label{sec:disper}

\begin{figure}[!htb]
\centering
\includegraphics[width=0.75\columnwidth]{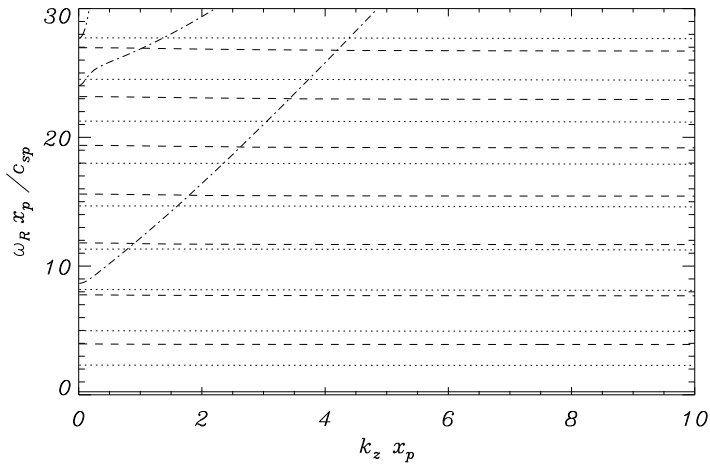}
\caption{Dimensionless real part of the frequency versus $k_z x_{\rm p}$ for the oscillatory symmetric modes: hybrid slow (solid line at the bottom), fundamental internal slow and first harmonics (dotted lines), fundamental external slow and first harmonics (dashed lines), fundamental internal fast and first harmonic (dash-dotted lines) and fundamental external fast (three dot-dashed line at the upper left corner).  \label{fig:kinkdisper}}
\end{figure}

Solutions of Eqs.~(\ref{eq:basic1})--(\ref{eq:basic3}) can be grouped in internal, external and hybrid modes. Although there is an infinite number of harmonics for internal and external modes, only two hybrid modes are possible: the hybrid slow mode and the hybrid fast mode \citep[this nomenclature is taken from][]{oliver}. Figure~\ref{fig:kinkdisper} shows the dimensionless real part of the frequency versus $k_z x_{\rm p}$ for the fundamental symmetric oscillatory modes (i.e. solutions with $v_x$ even with respect $x=0$) and some of their harmonics, where we have assumed the boundary conditions given by Eq.~(\ref{eq:set1}). A similar diagram can be obtained for the antisymmetric modes (i.e. solutions with $v_x$ odd with respect $x=0$) and for the other set of boundary conditions (Eq.~(\ref{eq:set2})).

\begin{figure}[!htb]
\centering
\includegraphics[width=1\columnwidth]{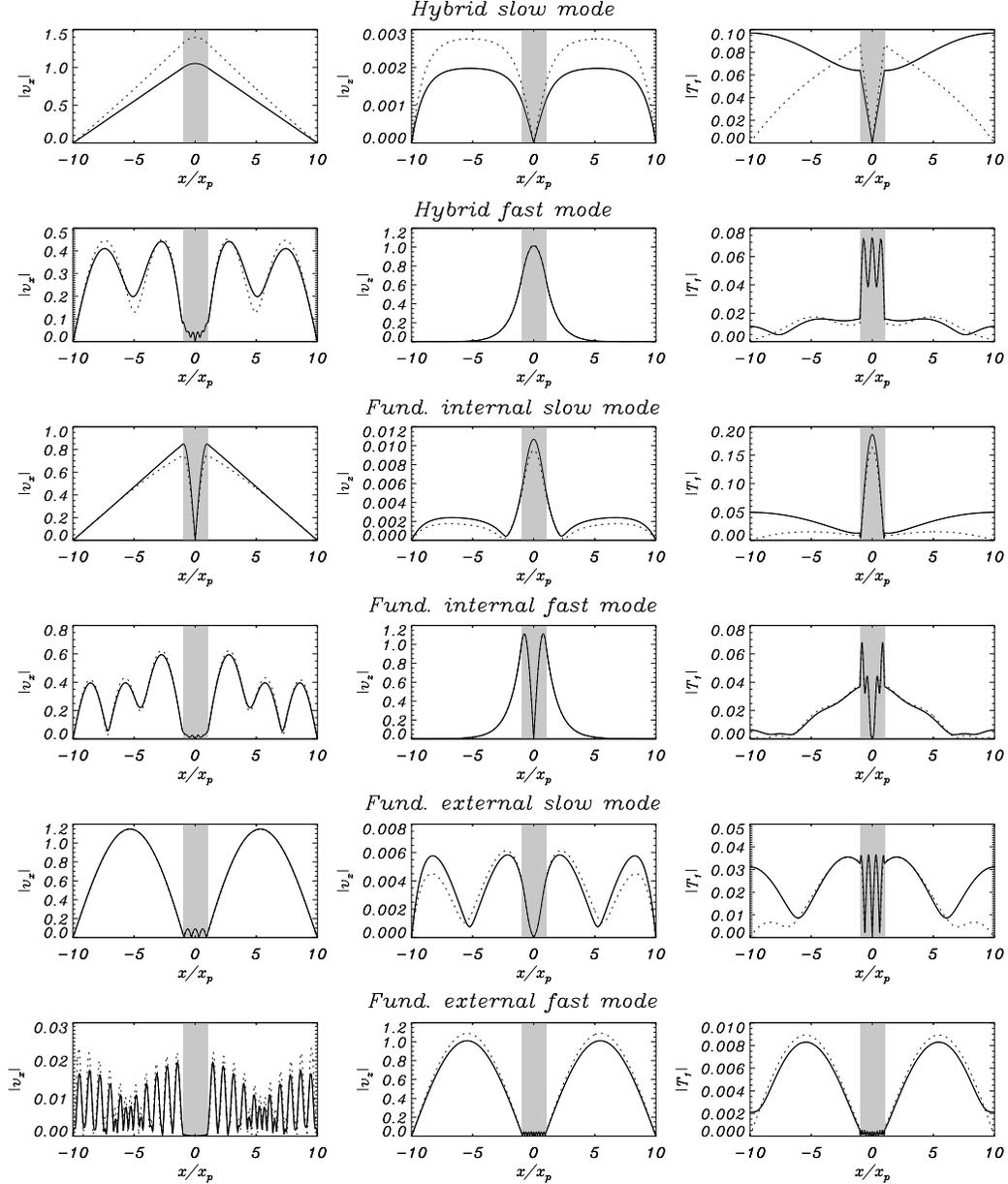}
\caption{Modulus of the eigenfunctions $v_x$, $v_z$ and $T_1$ (in arbitrary units) versus the dimensionless distance to the slab axis corresponding to the fundamental oscillatory modes for $k_z x_{\rm p} = 1$. The solid line corresponds to the boundary conditions $v_x = v_z = T_1' = 0$ at $x = \pm x_{\rm c}$, while the dotted line corresponds to the boundary conditions $v_x = v_z = T_1 = 0$ at $x = \pm x_{\rm c}$. The shaded region shows the location of the prominence slab. \label{fig:autofun}}
\end{figure}

The behaviour of the real part of the frequency is the same as that explained by \citet{oliver}. The value of $\omega_{\rm R}$ for both internal and external slow modes and for the hybrid slow mode shows a very weak dependence on $k_z$ since almost horizontal lines are seen in Fig.~\ref{fig:kinkdisper}. On the contrary, both internal and external fast modes show a quasi-parabolic dependence on $k_z$ (this is also applicable to the hybrid fast mode, present in the dispersion diagram for antisymmetric modes). The reader is referred to \citet{oliver} for more extensive details about the behaviour of $\omega_{\rm R}$.

Next, we focus on the fundamental modes and their eigenfunctions $v_x$, $v_z$ and $T_1$ are displayed in Fig.~\ref{fig:autofun} for $k_z x_{\rm p} = 1$ and for both sets of boundary conditions. The spatial structure of the disturbances $v_x$ and $v_z$ is the one shown by \citet{oliver}. Hence, non-adiabatic effects do not modify the spatial behaviour of velocity perturbations. Internal modes produce large plasma displacements inside the slab, external modes achieve large amplitudes in the corona and the amplitude of hybrid modes is of the same order in both media. It is worth to mention that the hybrid fast mode can be considered as an internal-like mode for large $k_z x_{\rm p}$ since the amplitude of its perturbations in the corona decreases as $k_z x_{\rm p}$ increases. Regarding the temperature perturbation, it is larger for the slow modes than for the fast modes and, in general, is larger in the prominence than in the corona. Finally, the differences between the eigenfunctions for the two sets of boundary conditions (Eqs.(\ref{eq:set1})--(\ref{eq:set2})) are only relevant for the hybrid slow mode.

From the observational point of view, internal and hybrid modes could be more easily observed than external modes by instruments focusing on prominences, since the amplitude of the latter ones is very small in the prominence body. For this reason, the results corresponding to internal and hybrid modes are the most interesting for prominence seismology. However, here we study the three kinds of solutions in order to perform a complete description of the fundamental wave modes supported by the equilibrium configuration.

\subsection{Mode coupling}

\citet{oliver} showed that avoided crossings occur in the dispersion diagram when two solutions couple and interchange their magnetoacoustic properties. Nevertheless, no avoided crossings seem to take place in our dispersion diagram (Fig.~\ref{fig:kinkdisper}) since the curves of $\omega_{\rm R}$ for the internal fast modes and for the slow modes cut each other. This fact can be understood by considering that in the present, non-adiabatic case the complete dispersion diagram is in a three-dimensional space because the frequency has an imaginary part. So, Fig.~\ref{fig:kinkdisper} actually corresponds to a projection of the complete three-dimensional dispersion diagram on the $k_z \omega_{\rm R}$--plane.

\begin{figure}[!htb]
\centering
\includegraphics[width=0.32\columnwidth]{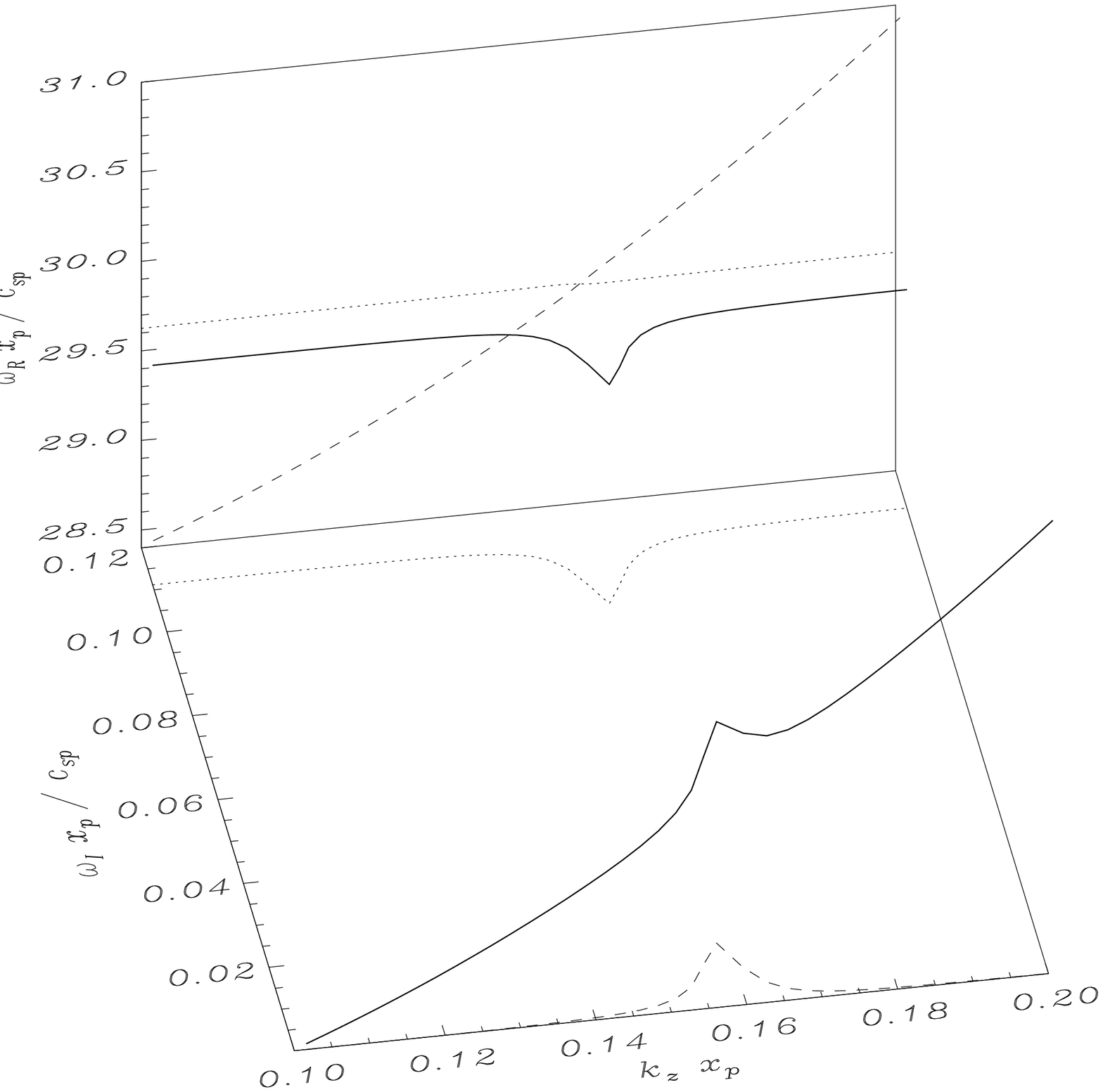}  
\includegraphics[width=0.32\columnwidth]{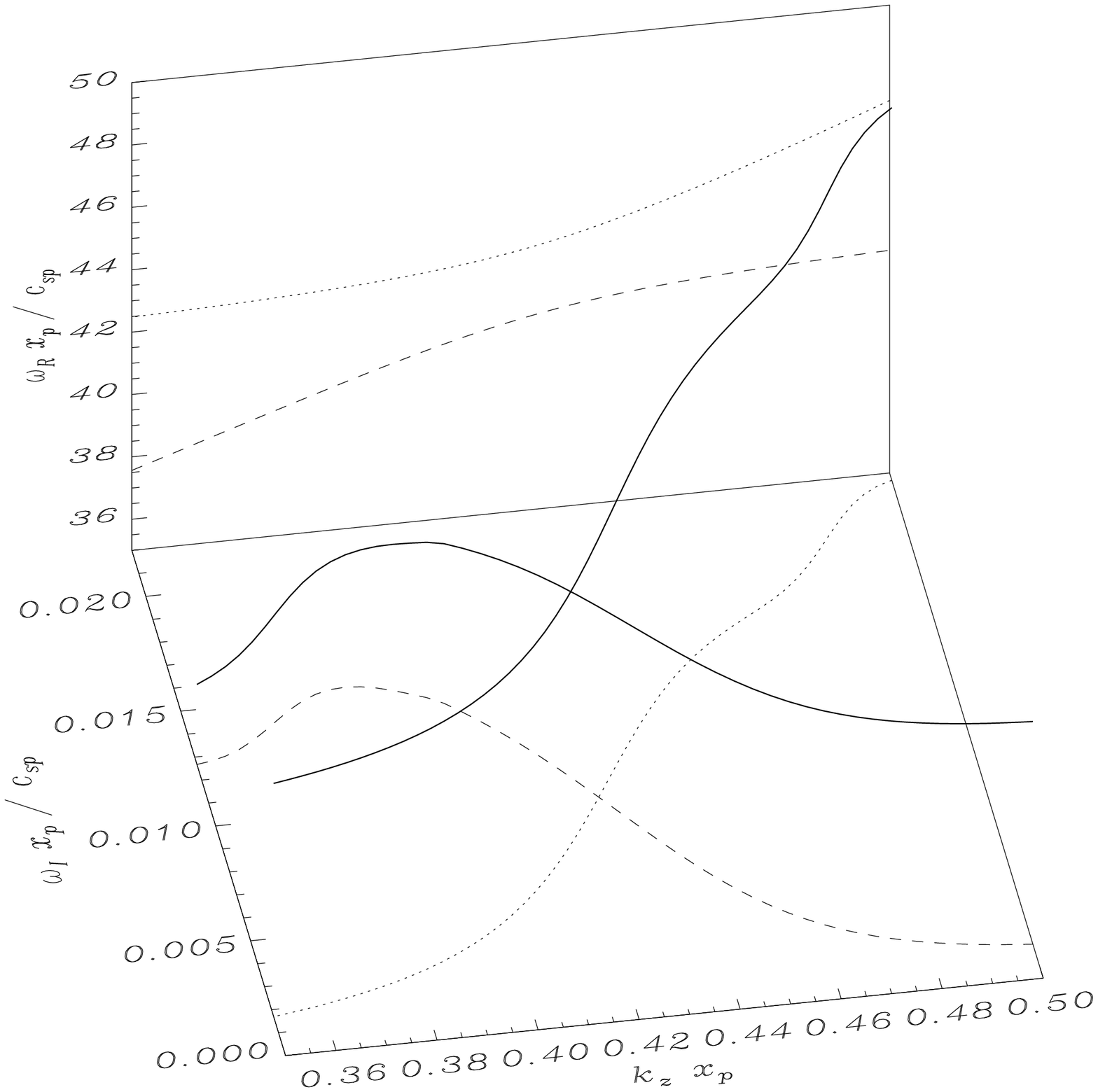} 
\includegraphics[width=0.32\columnwidth]{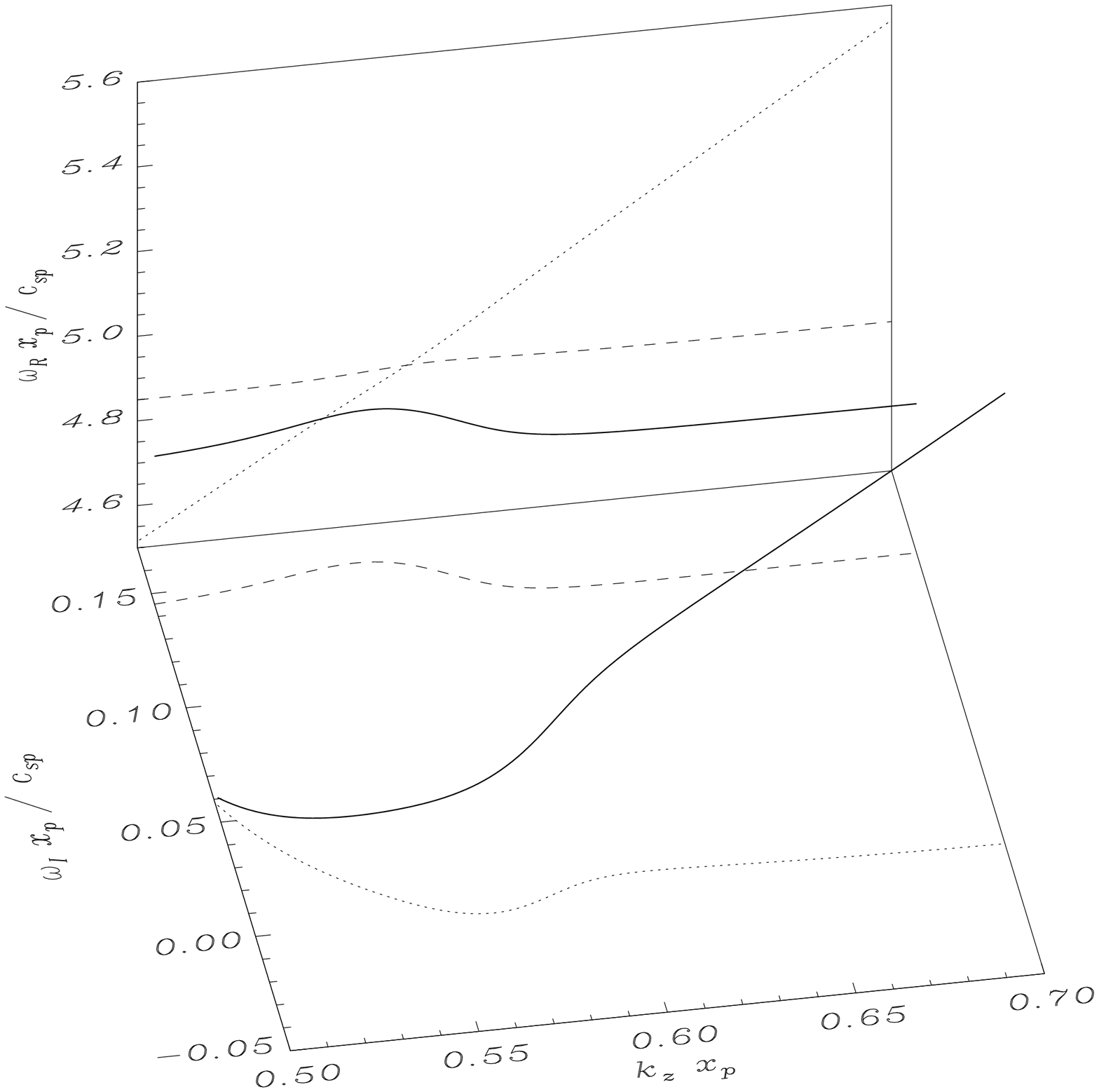} 
\caption{Three-dimensional dispersion diagrams (solid lines) close to a coupling between a fast mode and a slow mode. Dashed and dotted lines are the projections of the dispersion curves on the horizontal and vertical planes. The left-hand side panel presents a ``weak'' coupling, the middle panel shows a ``strong'' coupling and the right-hand side panel displays an ``anomalous'' coupling.  \label{fig:coupling}}
\end{figure}

Upon exploring the complete dispersion diagram, we have found that three different couplings can take place:
\begin{enumerate}
 \item If the imaginary parts of the frequency of the coupling modes differ by several orders of magnitude, there is no avoided crossing between the real parts. Hence, the coupling between modes is ``weak'' and only becomes apparent by means of a slight mutual approach of the imaginary parts of $\omega$ (see Fig.~\ref{fig:coupling}, left panel).
\item If both imaginary parts of the frequency have a similar value, the real parts show an avoided crossing and so a ``strong'' coupling takes place (see Fig.~\ref{fig:coupling}, mid panel).
\item In very peculiar cases, an ``anomalous'' coupling takes place when the imaginary parts of $\omega$ of the two coupling modes repel each other (see Fig.~\ref{fig:coupling}, right panel). This situation has important effects on the wave stability, as we explain in Sect.~\ref{sec:insta}. 
\end{enumerate}
The behaviour of the mode coupling was previously described by \citet{terradasrad01} in the cases that we call ``weak'' and ``strong'' couplings \citep[compare our Fig.~\ref{fig:coupling} with Fig.~12 of][]{terradasrad01}.

\subsection{Periods and damping times}
\label{sec:pertd}

\begin{figure}[!htb]
\centering
\includegraphics[width=1\columnwidth]{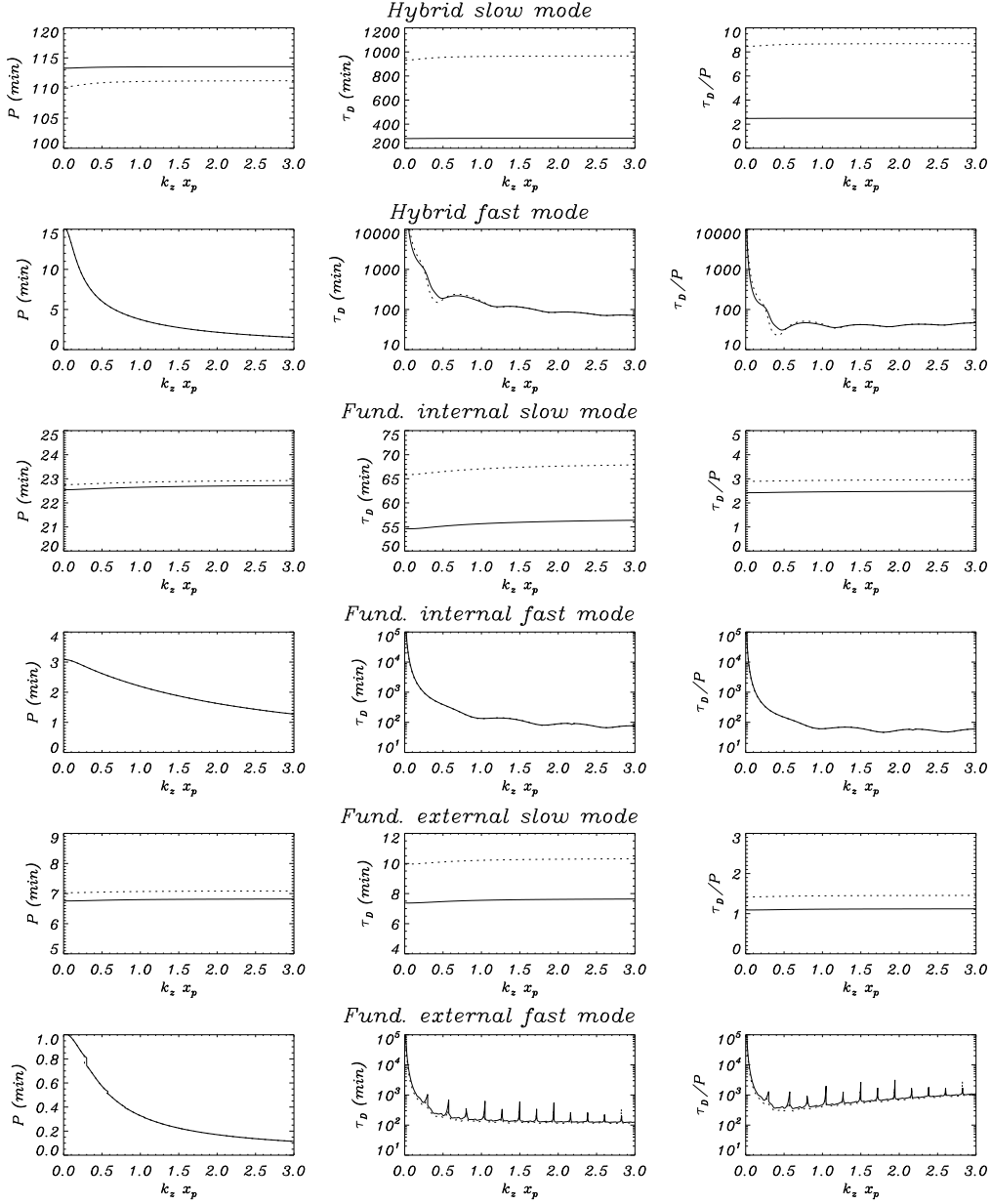}
\caption{Period (left), damping time (centre) and ratio of the damping time to the period (right) versus $k_z x_{\rm p}$ for the fundamental oscillatory modes. The solid line corresponds to the boundary conditions $v_x = v_z = T_1' = 0$, while the dotted line corresponds to the boundary conditions $v_x = v_z = T_1 = 0$.  \label{fig:kinkgen}}
\end{figure}

Hereafter we restrict ourselves to the fundamental modes and compute the oscillatory period, $P$, the damping time, $\tau_{\rm D}$, and the ratio of the damping time to the period as functions of the dimensionless wavenumber, $k_z x_{\rm p}$. We consider values for $k_z x_{\rm p}$ between 0.01 and 3, which correspond to wavelengths between $5\times 10^{3}\, \mathrm{km}$ and $10^{5}\, \mathrm{km}$, approximately. These values cover the range of typically observed wavelengths in prominence oscillations \citep{oliverballester02}. The results of the computations are displayed in Fig.~\ref{fig:kinkgen} considering the two sets of boundary conditions (Eqs.~(\ref{eq:set1})--(\ref{eq:set2})). 

The periods obtained here agree with those provided and commented by \citet{JR92} and \citet{oliver}, therefore we turn our attention to the damping times. Regarding slow modes, we see that they are strongly damped, with values of $\tau_{\rm D}/P$ close to 1 for the three modes. However, fast modes are much less attenuated and the obtained values of $\tau_{\rm D}/P$ are much larger than those observed. This fact involves an important difference with the results of Paper~I, in which fast waves were efficiently attenuated for some values of the wavenumber. Such as happens with the period, the damping time of slow modes is almost independent of $k_z$. On the contrary, fast modes are less attenuated for small $k_z$ than for large $k_z$. This evidence can be understood by means of the following arguments. Considering $k_z = 0$ then Eqs.~(\ref{eq:basic1})--(\ref{eq:basic3}) become,
\begin{equation}
 \cs^2 \frac{\der^2 v_x}{\der x^2} + \gamma \omega^2 v_x - \frac{i \omega \cs^2}{T_0} \frac{\der T_1}{\der x} = 0, \label{eq:basic1kz0}
\end{equation}
\begin{equation}
 \va^2 \frac{\der^2 v_z}{\der x^2} + \omega^2 v_z  = 0, \label{eq:basic2kz0}
\end{equation}
\begin{equation}
\kappa_\parallel \frac{1}{p_0} \frac{\der^2 T_1}{\der x^2} - \left( \omega_T + \frac{i \omega}{\gamma -1} \right) \frac{T_1}{T_0} - \left( 1+ \frac{i \omega_\rho}{\omega} \right) \frac{\der v_x}{\der x}  = 0. \label{eq:basic3kz0}
\end{equation}
Equations~(\ref{eq:basic1kz0}) and (\ref{eq:basic3kz0}) are still coupled and govern slow and thermal waves, which are affected by non-adiabatic mechanisms through the terms with $\kappa_\parallel$, $\omega_T$ and $\omega_\rho$ in Eq.~(\ref{eq:basic3kz0}). On the contrary, Eq.~(\ref{eq:basic2kz0}) is now decoupled from the rest and governs fast modes alone, which become pure Alfv\'en waves and are not affected by non-adiabatic terms. Thus, for $k_z \to 0$ fast waves tend to the ideal, undamped behaviour. When $k_z$ is increased, fast modes are more affected by acoustic effects and their damping time decreases and stabilises. The little peaks shown in the bottom panels of Fig.~\ref{fig:kinkgen}, corresponding to the external fast mode, are in fact the result of ``strong'' couplings with slow mode harmonics. The differences arising from the different boundary conditions are only of importance for the hybrid slow mode, as we indicated in Sect.~\ref{sec:disper}. We see that the boundary condition $T_1' = 0$ produces a substantially stronger damping for the hybrid slow mode than the condition $T_1 = 0$. 

Finally, an approximate value to the frequency of internal and external slow modes can be obtained by considering the approximation given in App.~B of Paper~I, namely
\begin{equation}
 \omega \approx \Lambda k_x, \label{eq:appslow}
\end{equation}
where $k_x$ is the wavenumber in the field direction and $\Lambda$ is the modified sound speed due to the presence of non-adiabatic effects, defined in Paper~I as follows
\begin{equation}
 \Lambda^2 \equiv \frac{\cs^2}{\gamma} \left[ \frac{\left( \gamma-1 \right) \left( \frac{T_0}{p_0} \kappa_\parallel k_x^2 + \omega_T - \omega_\rho \right) + i \gamma \omega}
{\left( \gamma -1 \right) \left( \frac{T_0}{p_0} \kappa_\parallel k_x^2 + \omega_T \right) + i \omega} \right]. \label{eq:Lambda}
\end{equation}
The value of $k_x$ is fixed by the equilibrium geometry, but for simplicity we consider now the analytical approximations of the dominant wavenumbers given by \citet{JR92} in the adiabatic case and for the long wavelength limit, namely
\begin{equation}
 k_x \approx \frac{\pi}{2 x_{\rm p}} \label{eq:kxint}
\end{equation}
for the fundamental internal mode, and
\begin{equation}
 k_x \approx \frac{\pi}{x_{\rm c} - x_{\rm p}} \label{eq:kxext}
\end{equation}
for the fundamental external mode. One must bear in mind that the wavenumbers in the present, non-adiabatic case are complex quantities, but we expect that their real part is similar to that in the adiabatic case, such as happens with the value of the frequency. Applying now Eq.~(\ref{eq:appslow}) to the internal slow mode, i.e. considering prominence parameters in the expression for $\Lambda$ (Eq.~\ref{eq:Lambda}) and the approximation for $k_x$ given by Eq.~(\ref{eq:kxint}), one obtains $P \approx$~23.50~min, $\tau_{\rm D} \approx$~73.05~min and  $\tau_{\rm D}/P \approx$~3.11. On the other hand, if the process is repeated for the external slow mode, this gives $P \approx$~6.20~min, $\tau_{\rm D} \approx$~6.70~min and $\tau_{\rm D}/P \approx$~1.08. We see that these approximate values reasonably agree with those numerically obtained and represented in Fig.~\ref{fig:kinkgen}.

\subsection{Importance of the damping mechanisms}
\label{sec:mecha}

\begin{figure}[!htb]
\centering
\includegraphics[width=\columnwidth]{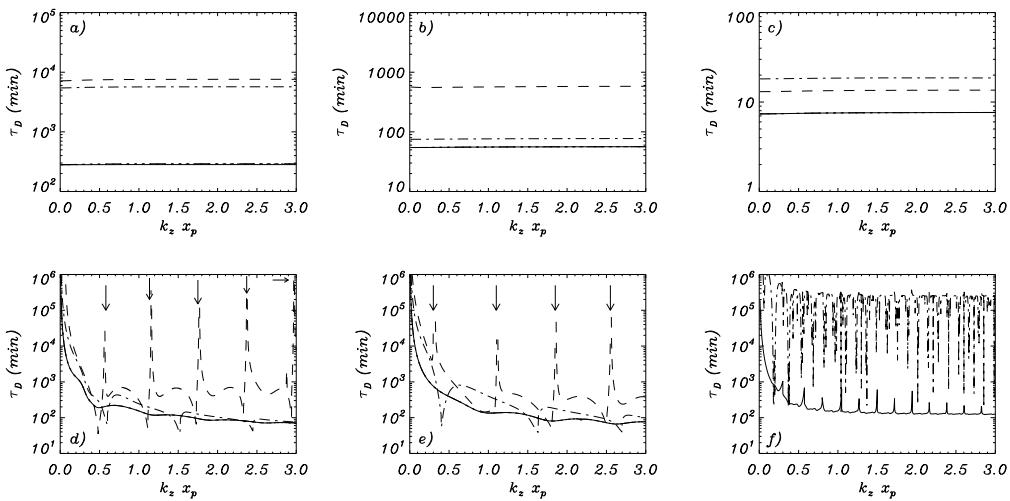}
\caption{Damping time versus $k_z x_{\rm p}$ for the fundamental modes: $a)$ hybrid slow,  $b)$ internal slow, $c)$ external slow, $d)$ hybrid fast, $e)$ internal fast, and $f)$ external fast. Different linestyles represent the omitted mechanism: all mechanisms considered (solid line), prominence conduction eliminated (dotted line), prominence radiation eliminated (dashed line), coronal conduction eliminated (dot-dashed line) and coronal radiation eliminated (three dot-dashed line). Arrows in panels $d$ and $e$ point the location of thermal instabilities ($\omega_{\rm I} < 0$) which appear if prominence radiation is omitted (dashed line). \label{fig:mechatot}}
\end{figure}

In order to know which are the mechanisms responsible for the damping of each mode, we now follow the same procedure as in Paper~I. We compare the damping time obtained when considering all non-adiabatic terms (displayed in the middle column of Fig.~\ref{fig:kinkgen}) with the results obtained when a specific mechanism is removed from the energy equation (Eq.~(\ref{eq:eneq})). This analysis allows us to know whether the omitted mechanism has a relevant effect on the attenuation. 

Before undertaking this investigation, we need to know if both sets of boundary conditions are adequate in the absence of thermal conduction. If one imposes $\kappa_\parallel = 0$ in Eq.~(\ref{eq:basic3}) then $T_1$ can be written as function of $v_z$ and $v_x'$,
\begin{equation}
 T_1 = - \frac{T_0 \left( 1 + i \omega_\rho / \omega \right) }{ \omega_T + i \omega / \left( \gamma - 1 \right) } \left(  v_x' + i k_z  v_z \right), \label{eq:temp}
\end{equation}
which can be substituted into Eqs.~(\ref{eq:basic1}) and (\ref{eq:basic2}) in order to obtain two coupled differential equations involving the perturbed velocities alone,
\begin{equation}
 \cs^2 \left( 1 + \frac{i \omega}{\hat{\omega}} \right) \frac{\der^2 v_x}{\der x^2} + \gamma  \omega^2 v_x + i k_z \cs^2 \left( 1 + \frac{ i \omega}{ \hat{\omega}} \right) \frac{\der v_z}{\der x} = 0, \label{eq:simply1}
\end{equation}
\begin{eqnarray}
  \va^2 \frac{\der^2 v_z}{\der x^2} &+& \left\{ \omega^2 - k_z^2 \left[ \va^2 + \frac{\cs^2}{\gamma} \left( 1 + \frac{ i \omega}{\hat{ \omega}}  \right) \right] \right\} v_z \nonumber \\ &+& i k_z \frac{\cs^2}{\gamma} \left( 1 + \frac{ i \omega}{\hat{ \omega}}  \right) \frac{\der v_x}{\der x} = 0. \label{eq:simply2}
\end{eqnarray}
Here $\hat{\omega}$ is introduced to simplify the notation,
\begin{displaymath}
 \hat{\omega} \equiv \frac{\omega_T + i \omega / \left( \gamma -1 \right)}{1 + i \omega_\rho / \omega}.
\end{displaymath}
Now, the system formed by Eqs.~(\ref{eq:simply1}) and (\ref{eq:simply2}) is fully determined by assuming only boundary conditions for $v_x$ and $v_z$. Hence, the behaviour of $T_1$ at the boundaries cannot be imposed but is fixed by the conditions over the velocity perturbations. If one takes $v_x = v_z = T_1' = 0$ as boundary conditions, then Eq.~(\ref{eq:temp}) yields the constraint $v_x'' + i k_z v_z' = 0$, which substituted in Eq.~(\ref{eq:simply1}) automatically gives the redundant condition $v_x = 0$. On the other hand, if one assumes $v_x = v_z = T_1 = 0$ at $x = \pm x_{\rm c}$, then Eq.~(\ref{eq:temp}) now imposes $v_x'  = 0$ at the boundaries. This last condition substituted in Eq.~(\ref{eq:simply2}) gives the extra condition $v_z'' = 0$ over the system, which implies a new restriction that is not generally satisfied by all solutions. Thus, $T_1' = 0$ reveals itself as the ``natural'' boundary condition for the temperature perturbation when thermal conduction is neglected. So, for the following investigation we restrict ourselves to the boundary conditions $v_x = v_z = T_1' = 0$ since the conditions $v_x = v_z = T_1 = 0$ are not consistent with the differential equations when thermal conduction is neglected.

 The results of the computations are displayed in Fig.~\ref{fig:mechatot}. Although we have explored a wide range of values of $k_z$, the plots are only drawn again for $ 0.01 < k_z x_{\rm p} < 3$ since we have found that the importance of the damping mechanisms does not show a strong dependence on $k_z$. Regarding slow modes, we clearly see that the damping of the internal mode is dominated by the radiation from the prominence plasma, as expected, while coronal conduction has a minor effect. On the other hand, the hybrid and external modes are affected by coronal conduction together with prominence radiation. Both mechanisms have a similar influence on the hybrid mode, while coronal conduction dominates the attenuation of the external mode. This result for the hybrid mode is coherent with the fact that its perturbations achieve large amplitudes both in the prominence and the corona (see top row of Fig.~\ref{fig:autofun}), so one expects that the most relevant damping mechanisms of each medium govern together the attenuation of the hybrid mode. However, the result for the external mode is \emph{a priori} surprising because its perturbations are very small in the prominence (see fifth row of Fig.~\ref{fig:autofun}) and one expects that the prominence-related mechanisms have a minor effect on its damping. The following discussion attempts to explain why prominence radiation affects so much the external mode.

The equilibrium configuration assumed in the present work implies an additional complication with respect to the equilibrium considered in Paper~I, in which magnetic field lines were taken parallel to the interface between the prominence and the corona. Hence, both media were thermally isolated in the model of Paper~I since there was no transfer of energy from one medium to the other. However, in the present model thermal conduction connects both media since field lines are transverse to the interfaces. This fact allows heat transfer between the prominence and the corona. So, some energy can flow along field lines and can be injected from the corona into the prominence, where the energy is efficiently radiated away by the plasma. In this way, the influence of prominence radiation on the damping of the external slow mode, and also the hybrid slow mode, is amplified by means of coronal thermal conduction.

Next we turn our attention to the fast modes. At first sight, the behaviour of the fast modes when a specific mechanism is removed from the energy equation is absolutely different from that seen in the case of the slow modes and needs more extensive explanations. In Sect.~\ref{sec:pertd}, we commented that the damping time of the fast modes is affected by the couplings with the slow modes. Now, we see that the nature of these couplings (being ``weak'', ``strong'' or ``anomalous'') changes depending on which is the non-adiabatic mechanism omitted in the energy equation. These changes in the coupling nature cause the damping time of the hybrid fast mode and the internal fast mode to vary from small values to very large values depending on the proximity to the couplings. So, we see that the consideration of both prominence radiation and coronal conduction has the effect of smoothing the curves of $\tau_{\rm D}$. 

In addition, the results corresponding to hybrid and internal fast modes show the appearance of thermal instabilities in very localised values of $k_z x_{\rm p}$ when prominence radiation is neglected (dashed lines), since then the interactions between fast modes and external slow modes leads to ``anomalous'' couplings. At these couplings, the value of $\omega_{\rm I}$ for the fast modes is pushed towards negative values (see the right-hand panel of Fig.~\ref{fig:coupling}). Such a situation has very important repercussions on the wave behaviour since for $\omega_{\rm I} < 0$ waves are amplified in time. The location of these instabilities in panels $d$ and $e$ of Fig.~\ref{fig:mechatot} have been pointed by means of arrows.

\subsection{Wave instabilities}
\label{sec:insta}

\begin{figure}[!htb]
\centering
\includegraphics[width=0.5\columnwidth]{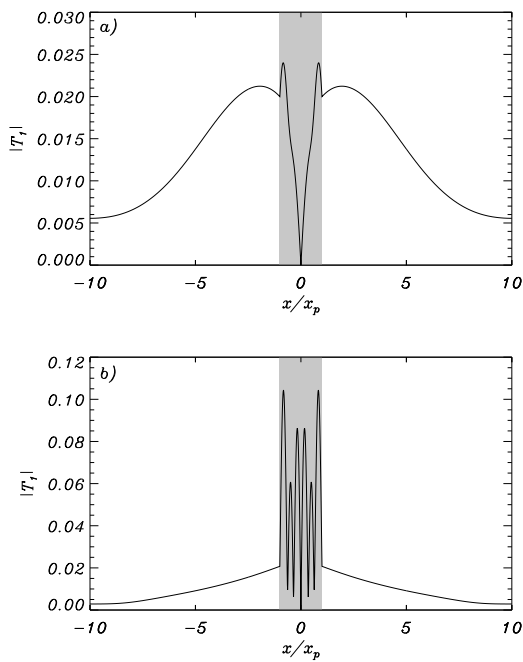}
\caption{Modulus of the eigenfunction $T_1$ (in arbitrary units) versus the dimensionless distance to the slab axis corresponding to the fundamental internal fast mode for $k_z x_{\rm p} \approx 0.3$ if $a)$ all non-adiabatic mechanisms are considered, and $b)$ without prominence radiation. The shaded region shows the location of the prominence slab. \label{fig:autoinst}}
\end{figure}

Wave instabilities discussed in Sect.~\ref{sec:mecha} require a more in-depth investigation. According to \citet{field}, the criterion for the appearance of wave instabilities is given by
\begin{equation}
\frac{\kappa_\parallel}{\rho_0} k_x^2 + L_T + \frac{1}{\gamma - 1} \frac{\rho_0}{T_0} L_\rho  < 0,  \label{eq:criterionfield}
\end{equation}
where $k_x$ is the wavenumber in the field direction. Results of \citet{carbonell}, see also Paper~I, point out that the heating scenario used in our calculations (constant heating per unit volume) cannot lead to thermal destabilisation. So, we can affirm that instabilities described in Sect.~\ref{sec:mecha} are not caused by the heating mechanism. In addition, instabilities only appear when radiative losses are omitted. In such situation, the instability criterion becomes
\begin{equation}
\frac{\kappa_\parallel}{\rho_0} k_x^2 < 0.  \label{eq:criterionfieldnorad}
\end{equation}
Equation~(\ref{eq:criterionfieldnorad}) is never satisfied unless an additional source of heating is present, which seems to be the present case. This extra energy source corresponds to heat injected from the corona into the prominence by thermal conduction, as was commented in Sect.~\ref{sec:mecha}. In the absence of radiation, prominence thermal conduction is the only mechanism that can dissipate this extra injected heat. One expects that in such situation the value of $k_x$ grows in order to increase the efficiency of prominence conduction. Figure~\ref{fig:autoinst} shows the eigenfunction of the temperature perturbation corresponding to the internal fast mode for $k_z x_{\rm p} \approx 0.3$ when all non-adiabatic mechanisms are considered, panel $a)$, and when radiative losses from the prominence plasma are omitted, panel $b)$. For this value of $k_z x_{\rm p}$, the wave becomes unstable ($\omega_{\rm I} < 0$) if prominence radiation is omitted. We see that smaller spatial-scales (i.e. larger $k_x$) are obtained within the prominence when prominence radiation is not taken into account, as expected. Although the efficiency of prominence conduction is increased in this way, it is still not enough to stabilise the perturbation.  

\begin{figure}[!htb]
\centering
\includegraphics[width=0.5\columnwidth]{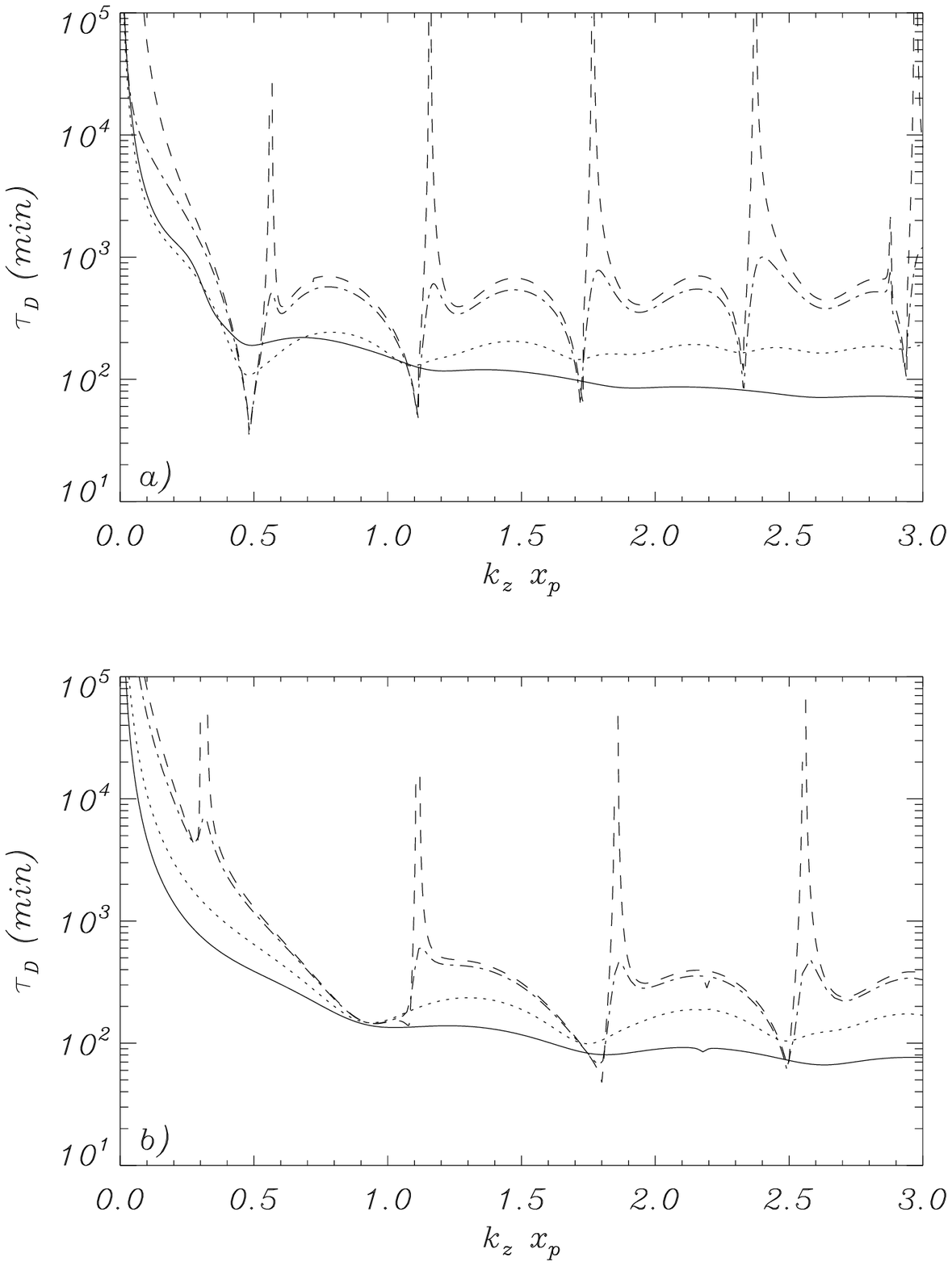}
\caption{Damping time versus $k_z x_{\rm p}$ for the fundamental $a)$ hybrid and $b)$ internal fast modes. The linestyles represent different optical thicknesses for the prominence plasma: Prominence (1) in solid line (this corresponds to the solid lines in Fig.~\ref{fig:mechatot}$d$ and $e$), Prominence~(2) in dotted line and Prominence~(3) in dot-dashed line. The dashed line corresponds to the results when the prominence radiation is omitted (dashed lines in Fig.~\ref{fig:mechatot}$d$ and $e$). The boundary conditions considered are $v_x=v_z=T_1'=0$. \label{fig:optical}}
\end{figure}

This last discussion points out that prominence radiative losses are of paramount importance to stabilise the disturbances. The efficiency of prominence radiation can be quantified by means of the radiation time-scale for the prominence plasma \citep{moortel},
\begin{equation}
 \tau_{\rm r} = \frac{\gamma p_0}{(\gamma -1) \rho^2_{\rm p} \chi_{\rm p}^* T_p^{\alpha_{\rm p}}}.
\end{equation}
Considering fixed equilibrium parameters, the value of $\tau_{\rm r}$ changes for different optical thicknesses of the prominence material (see regimes listed in Table~I of Paper~I). For Prominence~(1) parameters (optically thin plasma), $\tau_r \approx 309$~s, whereas for Prominence~(2) and Prominence~(3) regimes (optically thick and very thick plasma), $\tau_r \approx$~2,876~s and  $\tau_r \approx$~47,822~s, respectively, and so prominence radiation is less efficient. Obviously, $\tau_r \to \infty$ if the radiative term is omitted. The coronal plasma is always taken optically thin. Figure~\ref{fig:optical} shows the damping time of the fundamental hybrid and internal fast modes as a function of $k_z x_{\rm p}$ for the different prominence optical regimes. We see that the larger the optical thickness, the larger the damping time. This effect is especially relevant at the coupling points with the external slow modes, where thermal instabilities appear if radiative losses are completely inhibited.

\subsection{Exploring the parameter space}

In this Section we investigate how the attenuation of oscillations is affected by changing the equilibrium parameters. The motivation of this study is based on the fact that the estimated values for prominence plasma parameters, such as temperature, density, magnetic field strength or optical thickness, varies from one prominence to another, sometimes in a significant way \citep[e.g.][]{patsu}. Thus, it is important for our investigation to ascertain the sensitivity of the damping time to the equilibrium parameters around the values considered in our previous calculations. 

\begin{figure}[!htb]
\centering
\includegraphics[width=\columnwidth]{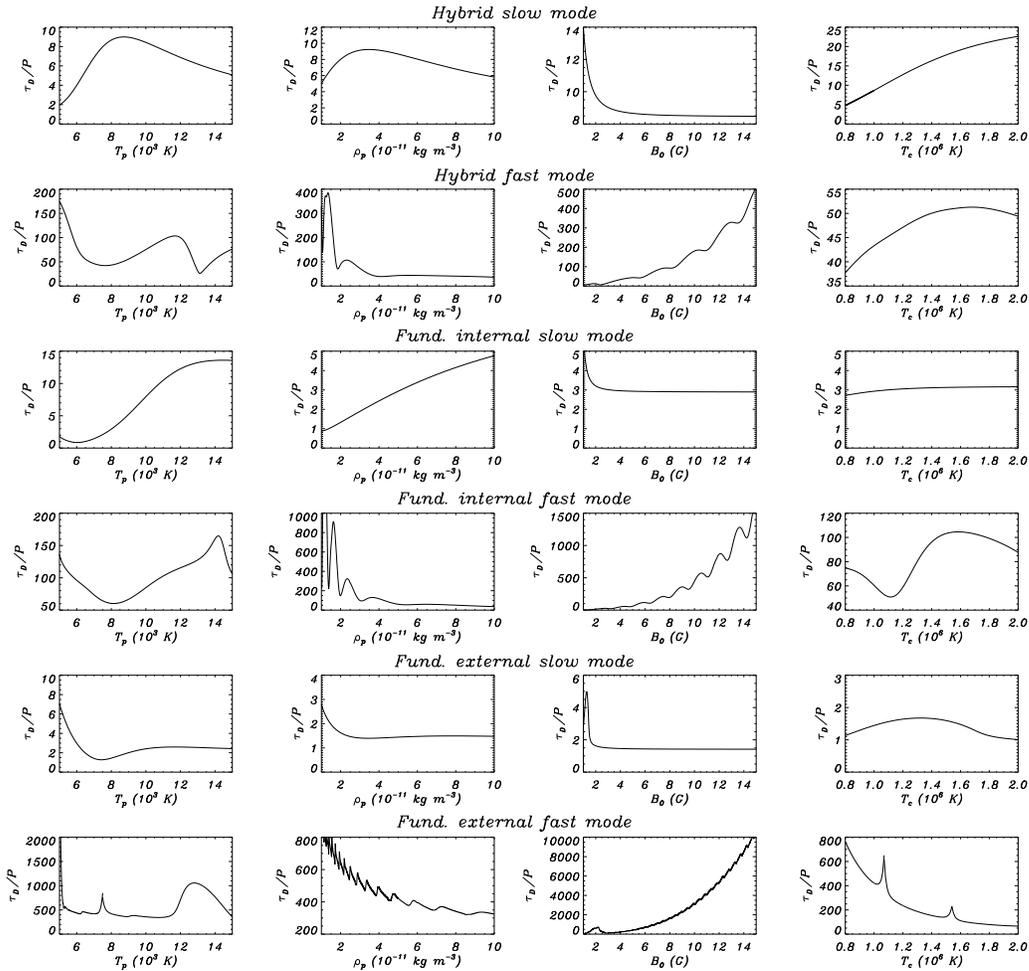}
\caption{Ratio of the damping time to the period for the fundamental oscillatory modes as function of, from the left to the right, the prominence temperature, the prominence density, the magnetic field strength and the coronal temperature. Computations performed considering  $k_z x_{\rm p} = 1$ and the boundary conditions $v_x=v_z=T_1=0$ at $x = \pm x_{\rm c}$. \label{fig:param}}
\end{figure}

First, we plot in Fig.~\ref{fig:param} the ratio of the damping time to the period corresponding to the fundamental modes as a function of equilibrium physical conditions, namely the prominence temperature, the prominence density, the magnetic field strength and the coronal temperature. The following ranges of values have been considered: 5000~K $< T_{\rm p} <$ 15,000~K; $10^{-11}$~kg~m$^{-3}$ $< \rho_{\rm p} <$ $10^{-10}$~kg~m$^{-3}$; 1~G ~$< B_0 <$ 15~G; and 800,000~K $< T_{\rm c} <$ 2,000,000~K. 

At first sight, we notice that the attenuation of fast modes is much more sensitive to the equilibrium conditions than the damping of slow modes. The attenuation of slow modes does not change in a significant way if the equilibrium physical conditions are modified, since the obtained $\tau_{\rm D} / P$ are always small and of the same order of magnitude. On the contrary, fast modes are highly sensitive especially to the prominence density and the magnetic field. It is noticeable that small values of $\tau_{\rm D} / P$ are obtained for the fast modes when large densities and weak magnetic fields are considered. If the magnetic field strength is increased or the prominence density is reduced, then $\tau_{\rm D} / P$ grows dramatically. Additionally, fast modes are again strongly affected by the couplings with slow modes, a fact that shows up in the form of very localised increases and decreases of $\tau_{\rm D} / P$.

On the other hand, we have studied the effect of considering a different heating scenario on the wave attenuation. In agreement with previous investigations \citep{carbonell,terradas}, results do not show significant discrepancies if different heating mechanisms are assumed.

Finally, we have also varied the length of magnetic field lines (by modifying the value of $x_{\rm c}$) and the prominence half-width, $x_{\rm p}$, in order to assess their effect on the damping time. For realistic values of both $x_{\rm c}$ and $x_{\rm p}$, no significant influences appear in the results with respect to those previously discussed. It is worth to mention that prominence conduction becomes a relevant mechanism for very a small, unrealistic prominence half-width ($x_{\rm p} \lesssim 10$~km), and coronal radiation is only important for very large, and again unrealistic, length of magnetic field lines ($x_{\rm c} \gtrsim 10^6$~km).

\subsection{Comparison with Terradas et al. (2005)}

\begin{figure}[!htp]
\centering
\includegraphics[width=\columnwidth]{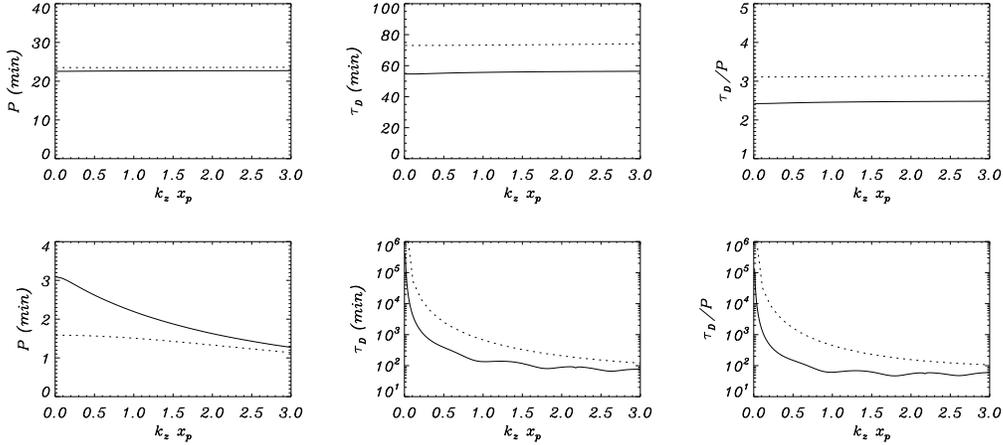}
\caption{Period (left), damping time (centre) and ratio of the damping time to the period (right) versus $k_z$ for the internal fundamental slow (upper panels) and fast (bottom panels) oscillatory modes. The solid lines are the solutions of a prominence plus corona equilibrium whereas the dotted lines represent the solutions of an isolated slab \citep{terradas}.  \label{fig:slab}}
\end{figure}

The final check of the importance of the coronal medium comes from the comparison between our results and those obtained by \citet{terradas} in the case of an isolated prominence slab (see Fig.~\ref{fig:slab}). Obviously, this comparison can only be performed for internal modes, since external and hybrid modes are not supported by an isolated slab. The boundary conditions assumed in the work of \citet{terradas} are $v_x = v_z = T_1' = 0$. According to the arguments given in Sect.~\ref{sec:mecha}, this condition for the perturbation to the temperature is the most suitable since thermal conduction is negligible in prominences. However, the line-tying condition at the edges of the prominence slab seems not to be the most appropriate election in the light of the eigenfunctions plotted in Fig.~\ref{fig:autofun}. Hence, our results point out that the interface between the prominence slab and the corona does not act as a rigid wall, and perturbations can be important in the corona even for internal modes.

Contrary to what was shown in Paper~I, in which only the fast mode was affected by the corona, in the present case both slow and fast modes of the isolated slab differ from those of a prominence--corona equilibrium. A decrement of the damping time is obtained for both waves in comparison with the solution of an isolated slab. The slow mode is less affected by the presence of the corona but the fast mode damping time is reduced by an order of magnitude, although it is still far from the observed values. As in the longitudinal magnetic field case, the consideration of the corona is of paramount importance for a correct description of the behaviour of oscillations and their attenuation, although its effect on the damping of fast modes is less noticeable than in the longitudinal case.

\section{Conclusions}
\label{sec:conclusions}

In this paper we have studied the wave attenuation in a system representing a quiescent solar prominence embedded in the coronal medium. The prominence has been modelled as a homogeneous plasma slab surrounded by a homogeneous medium with coronal conditions. Magnetic field lines have been assumed transverse to the prominence slab axis and the whole system has been bounded in the field direction by two photospheric rigid walls, in order to establish a realistic length for the field lines. The attenuation of the normal modes of such equilibrium has been investigated by considering parallel thermal conduction, radiative losses and plasma heating as non-adiabatic mechanisms, and focusing our study on the fundamental oscillatory modes. The main conclusions of this work are summarised next.
\begin{enumerate}
 \item Slow modes are strongly attenuated by non-adiabatic mechanisms, their damping times being of the order of the corresponding periods. Fast modes are less affected and present greater damping times.
\item The most relevant damping mechanisms are prominence radiation and coronal thermal conduction. The first one dominates the damping of internal modes, while the second one is responsible for the attenuation of external modes. The combined effect of both mechanisms governs the damping of hybrid modes. Neither prominence conduction nor coronal radiation become of importance for realistic values of the length of magnetic field lines and the prominence width.
\item The attenuation of slow modes is not affected by the value of the free component of the wavenumber, $k_z$. On the contrary, the behaviour of fast modes is strongly dependent on $k_z$.
\item Thermal conduction allows energy transfer between the prominence slab and the coronal medium. Prominence radiation has an essential role in dissipating the extra heat injected from the corona and stabilises the oscillations. Thermal instabilities appear if the radiative losses from the prominence plasma are omitted or significantly reduced (e.g. caused by an increase of the optical thickness) since the plasma cannot dissipate the extra injected heat in an efficient way.
\item The damping time of fast modes is strongly sensitive to the equilibrium physical parameters while slow waves are less affected by the variation of the equilibrium conditions.
\item The presence of the corona produces a decrement of the damping time of internal modes with respect to the solutions supported by an isolated prominence slab. Nevertheless, this effect is not enough to obtain damping times of the order of the period in the case of fast modes. 
\end{enumerate}

Considering the equilibrium parameters of Paper~I, the efficiency of non-adiabatic mechanisms on the damping of fast modes is smaller in the present case. This fact suggests that the orientation of magnetic field lines with respect to the slab axis has a relevant influence on the attenuation of fast modes, the configuration of Paper~I and the present one being limit cases. Moreover, fast modes are strongly sensitive to the equilibrium physical conditions, and it is possible to obtain small values of the damping time by considering extreme equilibrium parameters, such as very weak magnetic fields and very large prominence densities. In this way, fast modes show a wide range of theoretical damping times. On the other hand slow modes are always efficiently attenuated, with damping times of the order of their periods. This result suggests that the attenuation of prominence fast waves may be caused by other damping mechanisms not considered here. Some candidates could be resonant absorption \citep{arregui} and ion-neutral collisions \citep{forteza}. Among these mechanisms, resonant absortion may be a very efficient damping mechanism if non-uniform equilibria are considered, e.g. models with a transition region between the prominence and the corona. Other effects, as wave leakage, might only play a minor role in the damping of disturbances. Finally, future studies should take into account the prominence fine structure on the basis that small-amplitude oscillations are of local nature. Therefore, the investigation of the damping of fibril oscillations should be the next step.

     R.~O. and J.~L.~B. want to acknowledge the International Space Science Institute teams ``Coronal waves and Oscillations'' and ``Spectroscopy and Imaging of quiescent and eruptive solar prominences from space'' for useful discussions. The authors acknowledge the financial support received from the Spanish MCyT and the Conselleria d'Economia, Hisenda i Innovaci\'o of the CAIB under Grants No. AYA2006-07637 and PCTIB-2005GC3-03, respectively. Finally, R.~S. thanks the Conselleria d'Economia, Hisenda i Innovaci\'o for a fellowship.




\end{document}